\documentclass[preprint,prd,amsmath,amssymb]{revtex4}

\usepackage{graphicx}
\usepackage{dcolumn}
\usepackage{bm}

\usepackage{amsmath}
\usepackage{graphicx}
\usepackage{amssymb}
\usepackage{mathrsfs}
\usepackage{bbm}
\usepackage{epsfig}

\newcommand{\be}{\begin{equation}}
\newcommand{\ee}{\end{equation}}
\newcommand{\ba}{\begin{eqnarray}}
\newcommand{\ea}{\end{eqnarray}}
\newcommand{\bastar}{\begin{eqnarray*}}
\newcommand{\eastar}{\end{eqnarray*}}

\begin{document}
%
%

\renewcommand{\theequation}{\thesection.\\arabic{equation}}
\renewcommand{\thesection}{\arabic{section}}
\newcommand{\eqnull}{\setcounter{equation}{0}}

\numberwithin{equation}{section}

\vskip 1.0cm

\begin{titlepage}
\title{ {\bf  Knots And Swelling in Protein Folding } }

\author{Martin Lundgren}
\email{Martin.Lundgren@physics.uu.se}
\affiliation{Department of Physics and Astronomy,
Uppsala University,
P.O. Box 803, S-75108, Uppsala, Sweden}
\author{Antti J. Niemi}
\email{Antti.Niemi@physics.uu.se}
\affiliation{
Laboratoire de Mathematiques et Physique Theorique
CNRS UMR 6083, F\'ed\'eration Denis Poisson, Universit\'e de Tours,
Parc de Grandmont, F37200, Tours, France}
\affiliation{Department of Physics and Astronomy,
Uppsala University,
P.O. Box 803, S-75108, Uppsala, Sweden}


\vskip 5.0cm
\begin{abstract}
Proteins can sometimes be knotted, and for many reasons the 
study of knotted proteins is rapidly becoming very important. For example, 
it has been proposed that a knot increases the stability of a protein. Knots 
may also alter enzymatic activities and enhance binding. Moreover, 
knotted proteins may even have some substantial biomedical significance 
in relation to illnesses such as  Parkinson's  disease. But to a large extent the 
biological role of knots remains a conundrum. In particular, there is no explanation 
why knotted proteins are so scarce. Here we argue that knots are relatively rare 
because they tend to cause swelling in proteins that are too short, and presently short
proteins are over-represented in the Protein Data Bank (PDB). Using Monte Carlo
simulations we predict that the figure-8 knot leads to the most compact 
protein configuration when the number of amino acids is in the range of 
$200-600$. For the existence of the simplest knot, the trefoil, we estimate 
a theoretical upper bound of  $ 300-400$ amino acids, in line with the 
available PDB data.  
\end{abstract}

\maketitle
\end{titlepage}

\baselineskip .8cm
\renewcommand{\baselinestretch}{1}


\pagenumbering{arabic}
\renewcommand{\baselinestretch}{2}

\vskip 0.4cm

\section{introduction}

There are currently only some 300 known knotted 
proteins \cite{mans}-\cite{park} that are listed in the Protein Data Bank PDB \cite{pdb}. 
Furthermore, most of them correspond to the 
same protein but in a different species. Alternatively, they 
appear in multiple domain proteins, and often with the same knot 
repeated in each of the domains. When we only consider proteins  
in a single domain, there are no more than 17  different known 
knotted proteins \cite{mehran2}. With one single exception 
of a figure-8 knot, these are all trefoil knots 
with the value of central carbons $N$ in the range of $N \sim 82 -380$. Even
though there are examples of other knots such as the twist-3 knot, 
these have only been found in multiple domain proteins \cite{mehran2}.
In the present article  our goal is to identify some universal characteristics of
knotted proteins and to try and employ these to make predictions on the existence
of knots. 
In particular, we look for explanations why knots are so rare in the PDB
data.

Biologically active proteins are compact objects, and one
might expect that compactness is in some (yet unknown) 
manner important for their function. Consequently we propose that a study of the
relationship between knottedness and compactness could help to 
understand why knotted proteins are rare. Compactness can be measured 
by the Hausdorff dimension $d_H$ of the protein backbone,  
that can be determined from the scaling properties of the radius
of gyration $R_g$ \cite{huang}. In the limit where 
the number of central carbon atoms $N$ becomes very large $R_g$ 
obeys the scaling law
\be
R_g \ = \ \frac{1}{N+1} \sqrt{ \frac{1}{2} \sum_{i,j} ( {\bf r}_i 
- {\bf r}_j )^2 } \ \propto \ L \cdot N^{1/d_H}
\label{nu1}
\ee   
with ${\bf r}_i$ ($i=1,2,...,N$) the 
space coordinates of the central carbons.
Here $L$ is a dimensionfull swelling factor that 
sets the scale for the size of the protein, and
the inverse Hausdorff dimension $\nu = 1/d_H$ 
is called the compactness index.  
The swelling factor $L$ is not a universal quantity. But $\nu$ 
is universal: Different values of $\nu$ characterize 
different universality classes (phases) of proteins. 
Biologically active proteins that have $N$ in the range of 
$100 \leq N \leq 1.000$ obey the scaling law (\ref{nu1}) with 
$\nu \approx 0.378$ \cite{oma}. This is very close to the 
value $\nu = 1/3$ that determines the universality class of fully
collapsed protein (solid matter),
the difference is presumably due to some yet to be understood finite scaling 
effects \cite{binder}.

\vskip 0.4cm
\section{analysis of pdb data}

In Figure 1 we display the radius of gyration for the 17 known 
single domain knotted proteins that are presently listed in PDB, versus the number of their 
central carbons $N$. When we perform a least square linear fit 
to this data we find that the result is distorted by the 
shortest known knotted protein, the {\bf 2efv} for which N=82 \cite{pdb}. In order to
have a meaningful fit we proceed by leaving out {\bf 2efv}.
We also leave out {\bf 1ztu} since it is
(the only known) figure-8 protein. 
\begin{figure}
	\centering
	\includegraphics[width=0.60\textwidth]{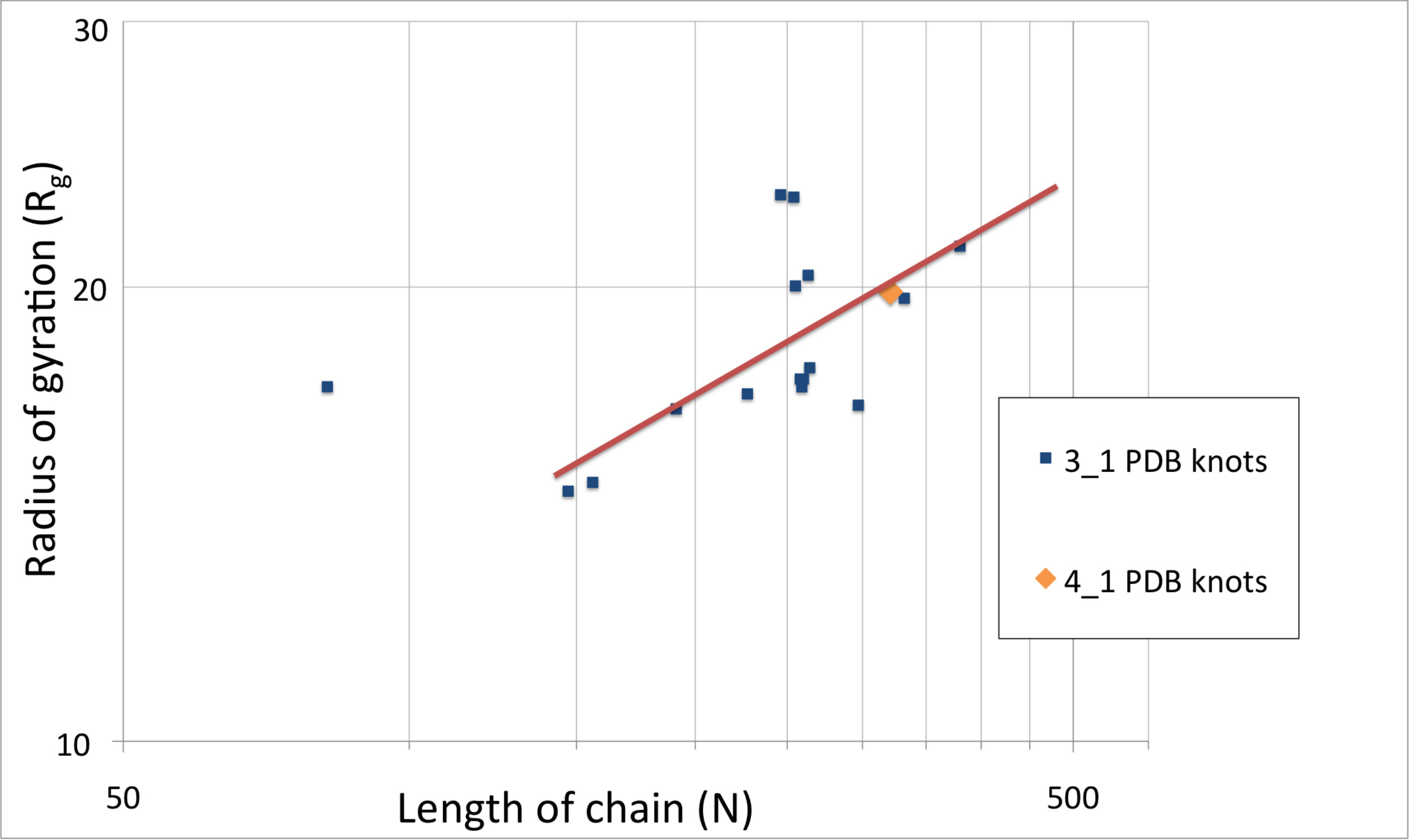}
\caption{\it Least square linear fit of radius of gyration
$R_g$ versus $N$ for the 15 single domain knotted 
proteins in PDB, as described in the text.
         }
        \label{Figure 1:}
\end{figure}

We are then left with the
15 trefoil proteins presently in PDB, and for these $N$ is in the range of $N \in [147 , 380]$. 
For these trefoils we obtain the following least square linear fit for the radius
of gyration,
\begin{equation}
R^{trefoil}_g \sim \ L \cdot N^\nu \ \sim \  (2.499 
\pm 0.661) \cdot N^{0.361 \pm 0.120}
\label{Rknot}
\end{equation}
We wish to compare this to the least square linear fit of $R_g$ to all
proteins presently in PDB. However, in order to ensure 
compatibility with (\ref{Rknot}), we choose only
those proteins for which $N> 125$. For these we get 
\begin{equation}
R^{all}_g \sim \ L \cdot N^\nu \ \sim \ (2.142 \pm 0.03) \cdot N^{0.387 \pm 0.003}
\label{Rall}
\end{equation}
but we note that if we include {\it all} proteins in PDB we obtain the estimate
\[
R^{all}_g \sim L \cdot N^\nu \ \sim \  (2.254 \pm 0.021) \cdot N^{ 0.378 \pm 0.002}
\]  
Despite slight differences
in the numerical values which is due to finite scaling effects, 
these two fits for $R^{all}_g$  are {\it very} close to
each other over the range of interest $N\in [ 125 , 400 ]$. In our analysis
we shall use both of them, and in this way we hope to 
control finite scaling corrections that are due to proteins with small values
of $N$. 

Since the quality of data that underlies (\ref{Rknot}) is poor,
one should be careful in drawing conclusions and with this in our 
mind we observe the following: 

Unknotted proteins have a clearly smaller 
value of $L$ than trefoils. Thus for small $N$ the unknotted proteins 
have a tendency to be more compact (smaller) than trefoil proteins. {\it When
size matters} this could explain why there are so few trefoils 
for small values of $N$ in the PDB data: The trefoil knot causes increased
swelling in small proteins.
 
But when $N$ grows, the swelling caused by the trefoil knot starts to
diminish. When $N$ reaches a value $N_c \approx 400 - 500$ {\it i.e.}
close to the upper bound $N = 400$ of our range, comparison between 
(\ref{Rknot}) and (\ref{Rall}) predicts that the trefoil proteins become 
equally compact than the unknotted ones. (The exact value of $N_c$
varies slightly depending on how we account for the finite scaling effects due to
short proteins.)  For $ N> N_c \approx 400-500$ our data for trefoil 
proteins is unreliable. But if the tendency continues there should 
be a range of values $N$ above $N_c$ where the presence of a 
trefoil improves compactness over proteins without knots. 
In this  range we expect that the relative number of trefoil
proteins increases. 

Note that asymptotically, for very large values of $N$, the scaling law
(\ref{nu1}) should be insensitive to the presence of a single localized knot.
\begin{figure}
	\centering
\includegraphics[width=0.60\textwidth]{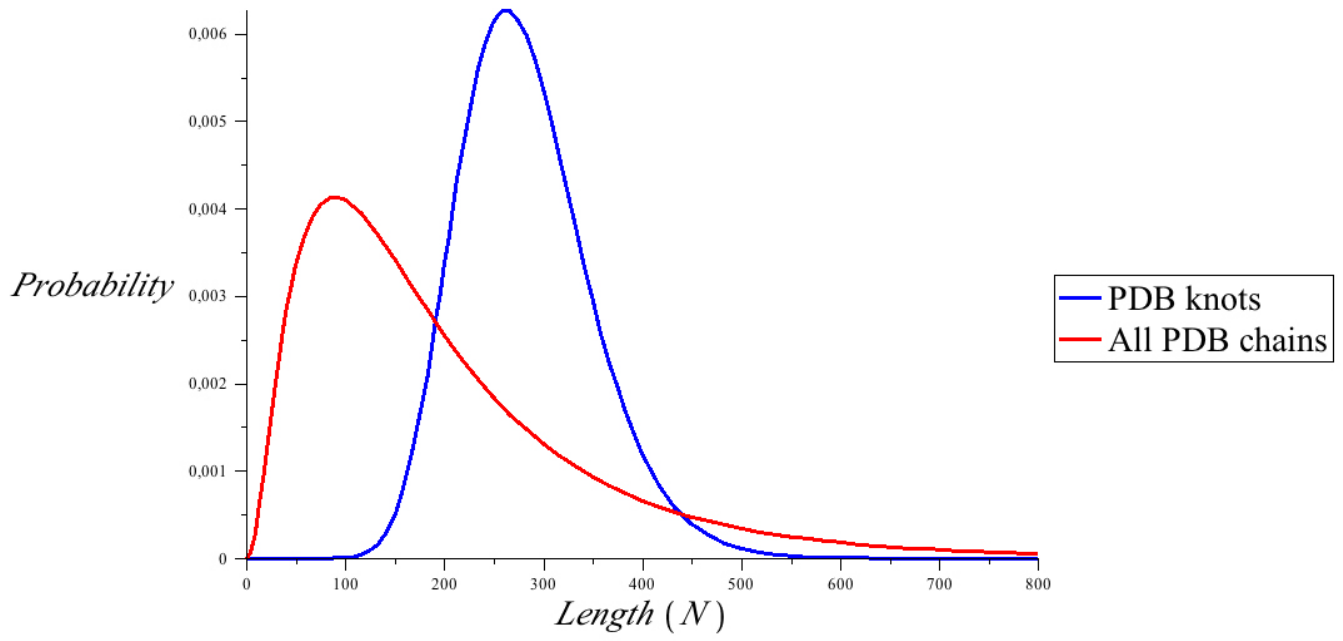}
\caption{\it Probability distributions of the generalized gamma 
form $p(N) \propto N^a \exp(-bN^c)$, fitted
to the number density of all proteins (red line) and  single domain trefoil 
proteins (blue line);  The latter are displayed in Figure 1.
         }
        \label{Figure 2:}
\end{figure}

In order to verify the reasonableness of the previous conclusion and in particular
whether there are additional reasons, we consider Figure 2 where 
we display the (probability) density function for the
number of proteins in PDB as a function of the length $N$, both for 
all proteins and for the single domain trefoil proteins. For 
all proteins the probability density peaks at around $N=90$ while for
trefoil proteins the peak is near $N=250$ where the total number of resolved
protein structures in PDB is already relatively small.
In particular, we observe that there are 
relatively few resolved protein structures with $N$ larger than $N_c \approx 400-500$ 
where our estimates predict that
trefoil proteins might become less swollen than unknotted ones. Consequently the
small number of trefoil knots could be partly due to the scarcity of data in PDB.
This is in line with our observation on the behaviour of (\ref{Rknot}), (\ref{Rall})
for $N > 400$. 

In addition, since the probability density of trefoils is sharply peaked
within the range $N \approx 150-400$ a partial explanation could also be that 
there is some yet unknown reason why proteins with trefoil knots prefer these 
values of central carbons.

\vskip 0.4cm
\section{theoretical estimates}

In order to better understand the effect of knots on protein swelling, and 
in particular to clarify whether the presence of trefoil knots 
tends to decrease or increase the compactness of proteins when $N>400$ 
we have theoretically investigated
how different knots influence the radius of gyration of native state 
proteins when $N$ is within the range $125 \leq N \leq 800$. For this
we have employed the Landau-Ginsburg model of protein folding that we have described
in \cite{oma}. Since the model provides a good description of 
the universal aspects of protein folding in particular for the 
mostly-$\alpha$ and $\alpha/\beta$ family of proteins that have been found
to support knots \cite{mehran2}, we expect it to have good predictive power
on the effects of knots on protein swelling.

We have performed extensive Monte Carlo simulations with two different sets 
of knotted proteins. We summarize the results in Table 1.
The {\it first set} consists of chains that have 
either one, three or five distinct trefoil knots along the backbone. 
The {\it second set} consists of chains where the proteins have either 
the trefoil knot $(3_1)$, the figure-8 knot $(4_1)$, or the twist-3 knot $(5_2)$ 
along their backbone. The simulations have been performed by selecting up
to 10 different values of $N$ in the ranges dispalyed in Table 1,
with the exact values depending on the knot 
complexity (see Table 1), and then performing around 80 independent runs 
at each of these value $N$ to compute the radius of gyration; See Figure 3. 
\begin{figure}
	\centering
\includegraphics[width=0.60\textwidth]{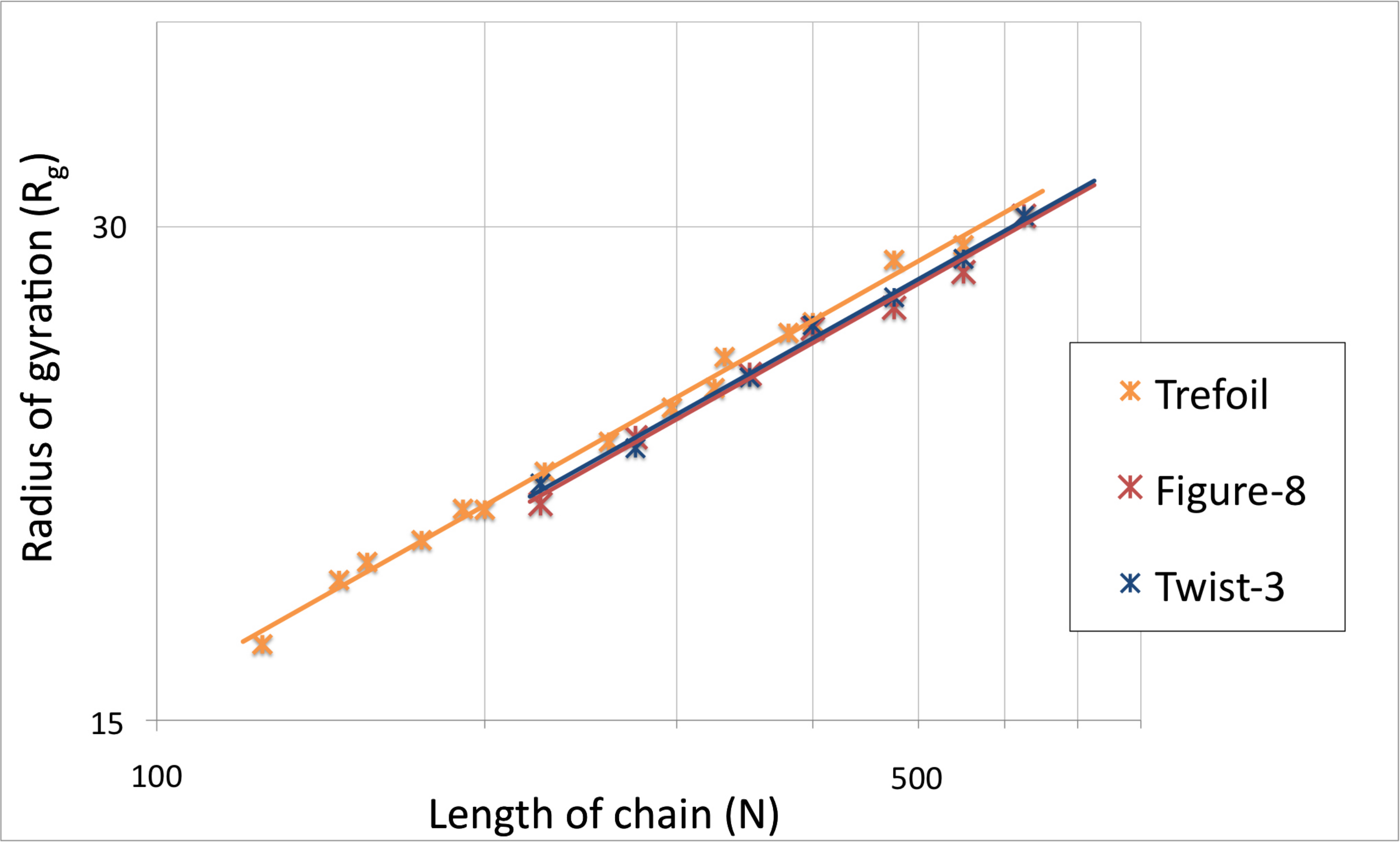}
\caption{\it Least square linear fits of $R_g$ for the trefoil ($3_1$),
figure-8 ($4_1$) and twist-3 ($5_2$) proteins. }
        \label{Figure 3:}
\end{figure}
The initial 
configuration is a (relatively tight) knot 
located deep inside the protein structure. In each of the cases 
we find that the least square linear fit of (\ref{nu1}) provides an excellent 
match for the data, we have not been able to identify any kind of systematic 
nonlinear corrections.
\begin{table*}[htbp]
	\centering
\begin{tabular}{|c|c|c|c|c|c|c|}
\hline 
knot type & $L$ & $\nu$ & $\Delta L$ & $\Delta \nu$ & $N_{knot}$ & $N_{range}$ \tabularnewline
\hline \hline
single  $3_1$ & 2.719 & 0.378 & $\pm$0.072 & $\pm$0.013 & 50-70 & 125-550 \tabularnewline
\hline 
three $3_1$     & 2.552 & 0.395 & $\pm$0.109 & $\pm$0.019 & & 225-550 \tabularnewline
\hline 
five $3_1$   & 2.500 & 0.403 & $\pm$0.167 & $\pm$0.027 & & 350-800 \tabularnewline
\hline 
$4_1$         & 2.729 & 0.373 & $\pm$0.109 & $\pm$0.018& 70-100 & 225-625 \tabularnewline
\hline 
$5_2$         & 2.764 & 0.372 & $\pm$0.084 & $\pm$0.014 & 90-110 & 225-625 \tabularnewline
\hline 
\end{tabular}
\caption{Swelling factor $L$ and compactness index $\nu$, with corresponding
standard errors $\Delta L$ and $\Delta \nu$ and an estimate for the average 
length of knots in the number of central carbons $N_{knot}$,
for different knot types simulated using the model described in \cite{oma}. 
The last column gives the range of values of $N$ for which the
simulations have been performed. The lower bound is
selected to accomodate a deep knot.}
	\label{tab:knots1}
\end{table*}
\noindent
But in order to eliminate the influence of finite scaling effects that are due to short
unknotted backbones, following (\ref{Rknot}), (\ref{Rall}) we compare the 
knotted proteins to unknotted ones using a set of different ranges of values
$N$ for the latter. These are given in Table 2.
\begin{table*}[htbp]
	\centering
\begin{tabular}{|c|c|c|c|c|}
\hline 
 range of $N$ & $L$ & $\nu$ & $\Delta L$ & $\Delta \nu$  \tabularnewline
\hline\hline
$75<N<1.000$ & 2.656 & 0.379 & $\pm$0.049 & $\pm$0.008 \tabularnewline
\hline 
$125 < N < 1.000$  & 3.103 & 0.353 & $\pm$0.079 & $\pm$0.013  \tabularnewline
\hline 
$175 < N < 1.000$ & 3.411 & 0.338 & $\pm$0.093 & $\pm$0.015  \tabularnewline
\hline 
$250 < N < 1.000$ & 3.522 & 0.333 & $\pm$0.128 & $\pm$0.02  \tabularnewline
\hline 
\end{tabular}
\caption{Least square linear fits for $L$ and $\nu$ 
with standard errors $\Delta L$ and $\Delta \nu$, computed for 
unknotted proteins with varying range for $N$.}	\label{tab:knots2}
\end{table*}

We observe from Table 2 that the small $N$ finite scaling corrections 
tend to systematically decrease the value of $L$ and increase the value 
of $\nu$. Already for the range $250 < N < 1.000$ we find that $\nu$ is 
very close to its theoretical value $\nu = 1/3$ corresponding to totally 
collapsed proteins.

 In the case of the {\it first set} we compare the unknotted proteins with 
proteins that have either one, three or five trefoil knots along their backbone.

For proteins with a single trefoil ($3_1$), we find that there is a tendency
for long trefoil proteins to be more {\it swollen} than the unknotted ones.
We estimate that a transition occurs at around $N_c \approx 300$, and for
$N < N_c$ our simulations predict that trefoils are (slightly) more compact
than unknots. Since we estimate that there is a minimum length of about 50-70 
central carbons for a trefoil knot to form, our simulations suggest that deep
trefoil knots should predominantly be present for values $N 
\approx 100 - 300$. This conclusion is in line with the PDB data in Figure 1 over
its range of validity, and consistent with the probability density 
displayed in Figure 2.
 
We find that the presence of several trefoils along the backbone clearly
{\it increases} swelling for all values of $N$ we have studied. 
Consequently our simulations suggest that these configurations
should be very rare.
Indeed, in PDB data multiple trefoil knots have until now only been observed 
in multiple domain proteins, with only a single knot in the independent domains. 

In the case of our {\it second set}  
we compare the unknotted proteins to proteins with two more complex
knots, the figure-8 ($4_1$) knot and the twist-3 ($5_2$) knot along the 
protein backbone. 

For the figure-8  
knot in our range of $N$ we find that the ensuing proteins are slighty {\it more compact}
than the unknotted ones. We find an upper bound $N_c \approx 600$ beyond
which the figure-8 proteins start to become more swollen than the unknotted
ones. Since we estimate that the lower length of a figure-8 knot is close
to 100 central carbons, we propose that these knots should be present in PDB
data as deep knots, predominantly with $N$ in the range between 150 
and 600. Thus far only one has been found, {\bf 1ztu} with $N=320$ \cite{mehran2}. But 
as visible in Figure 2 there are also
relatively very few protein structures that have been resolved within this range of $N$.

In the case of the twist-3 knot, we find that proteins with this
knot are slightly more
swollen than the figure-8 knots. But they appear more compact than unknots
at least until $N$ reaches a value $N_c \approx 500$ beyond which they appear
to swell more than unknots. Furthermore, our estimates suggest 
that there is a lower bound at around $N \approx 300$ below which
these protein knots begin to be more swollen than unknots. However, this estimate  
is to some extent plagued by finite scaling effects due to the presence of short unknotted 
proteins in the analysis. We note that the two 
known $5_2$ knots \cite{mehran1} appear in multiple domain proteins, one ({\bf 1xd3}) 
in a domain with $N=228$ and the other
({\bf 2etl}) in a domain with $N=223$ central carbons.

Finally, in the theoretically important $N \to \infty$ limit 
the effect of a (localized) 
knot to the scaling law (\ref{nu1}) must vanish: Thus the presence of a knot leads
to an intricate non-linear finite scaling effect that remains to be understood in detail.

\vskip 0.4cm
\section{conclusions}
In summary, knots are examples of topologically nontrivial structures in 
proteins with potentially very high biomedical relevance. However, until now 
knots have been identified in only relatively few single domain proteins.
In order to understand the reason we have analysed the data in PDB to conclude, 
that in the case of short proteins the (trefoil) knots have a 
tendency to increase the swelling of the folded protein backbone. 
But when the backbone length increases, the swelling due to trefoil
seems to decrease and from the PDB data we estimate that for 
backbones with $N \approx 400-500$ central carbons 
the swelling due to the trefoil disappears. 
The reason why knots are so rare in PDB data would then be 
partially explained by the fact that until now the structure of 
only relatively short proteins have been reliably resolved. 

But in addition,  
it could be that for some reason trefoil knots only 
appear for $ N \approx 100 - 400$. In order to resolve
this and clarify what effect knots have on protein swelling, we have
performed extensive Monte-Carlo simulations using a Landau-Ginsburg 
model of protein folding. We find that for trefoil knots the 
swelling is minimal when $N$ is within the range of $100-300$.
But in contrast to the interpolated PDB data, beyond $N_c \approx 300$ we predict 
that trefoils tend to increase swelling. This prediction is consistent 
with Figure 2 that shows a clear peak of trefoils near $N \approx 250$. 
In the case of multiple trefoil knots, we find increased swelling in 
all cases and consequently these configurations
should remain quite rare unless the protein chains become much longer.  

But when we increase the knot complexity, we find that it leads to an improved 
compactness for a range of values of $N$.  We note that by using a very different
method, the authors of \cite{vkk} arrived at a very similar conclusion.
The effect appears to
be most profound for figure-8 knot for which we predict an eventual relative 
increase in their number in PDB data as long as $N$ is less than
$\sim 600$ but large enough to support a deep knot. For twist-3 we predict
an eventual relative increase in their number in PDB data, until $N$ 
reaches a value $\sim 500$. However, since
the twist-3 appears to be (slightly) more swollen 
than figure-8, it should remain more rare. Since the numerical differences in swelling
that are revealed in our simulations are quite 
smal andl our conclusions are based more on tendencies than clear numerical
differences, the effects of various biological and 
evolutionary factors may eventually turn out to be more dominant
than swelling. More exhaustive numerical simulations that in particular 
take into account the detailed amino acid structures of the proteins are thus
needed before more detailed
predictions on swelling can be made. Furthermore, it remains a theoretical 
challenge to understand the universal structure of finite scaling corrections to the 
radius of gyration in the case of collapsed protein chains.

\vskip 0.4cm
\section{acknowledgements}
Our research is supported by grant from the Swedish Research Council (VR). 
We both thank U. Danielsson for many discussions during
this work, and for commenting the manuscript. A.J.N. thanks M. Kardar and 
in particular P. Virnau for communications, and M. Luescher for a discussion.
A.J.N. thanks CERN for hospitality during this work.

\vfill\eject

\end{document}